\documentclass[useAMS,usenatbib]{mn2e}
\usepackage{graphicx,amssymb,epsfig}
\usepackage{natbib}
\def\msun{\,{\rm M_\odot}}
\def\simlt{\mathrel{\rlap{\lower 3pt\hbox{$\sim$}}\raise 2.0pt\hbox{$<$}}}
\def\simgt{\mathrel{\rlap{\lower 3pt\hbox{$\sim$}} \raise 2.0pt\hbox{$>$}}}
\def\lta{\mathrel{\rlap{\lower 3pt\hbox{$\sim$}}\raise 2.0pt\hbox{$<$}}}
\def\gta{\mathrel{\rlap{\lower 3pt\hbox{$\sim$}} \raise 2.0pt\hbox{$>$}}}
\title[Evolution of massive black hole seeds]
{The evolution of  massive black hole seeds}

\author[Volonteri, Lodato \& Natarajan]{Marta Volonteri$^{1}$, Giuseppe Lodato$^{2}$
\& Priyamvada Natarajan$^{3,4}$\\
\footnotemark[1] 
$^{1}$ Department Of Astronomy, University of Michigan, Ann Arbor, MI, USA\\
$^{2}$ Department of Physics and Astronomy, University of Leicester, Leicester,
LE1 7RH, UK\\
$^{3}$Department of Astronomy, Yale University, P. O. Box 208101, 
New Haven, CT 06511-208101, USA \\
$^{4}$Department of Physics, Yale University, P. O. Box 208120, 
New Haven, CT 06520-208120, USA}

\begin{document}

\maketitle

\begin{abstract}
We investigate the evolution of high redshift seed black hole masses
at late times and their observational signatures.  The massive black
hole seeds studied here form at extremely high redshifts from the
direct collapse of pre-galactic gas discs. Populating dark matter
halos with seeds formed in this way, we follow the mass assembly of
these black holes to the present time using a Monte-Carlo merger
tree. Using this machinery we predict the black hole mass function at
high redshifts and at the present time; the integrated mass density of
black holes and the luminosity function of accreting black holes as a
function of redshift. These predictions are made for a set of three
seed models with varying black hole formation efficiency. Given the
accuracy of current observational constraints, all 3 models can be
adequately fit. Discrimination between the models appears
predominantly at the low mass end of the present day black hole mass
function which is not observationally well constrained. However, all
our models predict that low surface brightness, bulgeless galaxies
with large discs are least likely to be sites for the formation of
massive seed black holes at high redshifts. The efficiency of seed
formation at high redshifts has a direct influence on the black hole
occupation fraction in galaxies at $z=0$. This effect is more
pronounced for low mass galaxies.  This is the key discriminant
between the models studied here and the Population III remnant seed
model.  We find that there exists a population of low mass galaxies
that do not host nuclear black holes. Our prediction of the shape of
the $M_{\rm bh} - \sigma$ relation at the low mass end is in agreement
with the recent observational determination from the census of low
mass galaxies in the Virgo cluster.
\end{abstract}
\begin{keywords}

\end{keywords}

\section{Introduction}

The demography of local galaxies suggests that most galaxies host a
quiescent supermassive black hole (SMBH) at the present time and the
properties of the black hole are correlated with those of the host
spheroid.  In particular, recent observational evidence points to the
existence of a tight correlation between the mass of the central black
hole and the velocity dispersion of the host spheroid
\citep{tremaine02,ferrarese2000,gebhardt2000b} in nearby
galaxies. This correlation strongly suggests coeval growth of the
black hole and the stellar component via likely regulation of the gas
supply in galactic nuclei
\citep{silk98,kauffmann00,king03,thompson05}.

Black hole growth is believed to be powered by gas accretion
 \citep{lyndenbell69} and accreting black holes are detected as optically
 bright quasars. These optically bright quasars appear
 to exist out to the highest redshifts probed at the present time. 
Therefore, the mass build-up of SMBHs
 is likely to have commenced at extremely high redshifts ($z > 10$). In fact,
 optically bright quasars have now been detected at $z > 6$ (e.g.,
 \citealt{fan04,fan06}) in the Sloan Digital Sky Survey (SDSS). 
Hosts of high redshift quasars 
 are often strong sources of dust emission \citep{omont01,cox02,carilli02,walter03},
 suggesting that quasars were already in place in massive galaxies at a time
 when galaxies were undergoing vigourous star formation. The growth
 spurts of SMBHs are also seen in the X-ray waveband. The integrated
 emission from these X-ray quasars generates the cosmic X-ray
 background (XRB), and its spectrum suggests that most black-hole
 growth is obscured in optical wavelengths
 \citep{fabian99,mushot00,hasinger01,barger03,barger05,worsley05}. There
 exist examples of obscured black-hole growth in the form of
 `Type-2' quasars, but their detected numbers are fewer than expected
 from models of the XRB. However, there is recent tantalizing evidence from
 infra-red (IR) studies that dust-obscured accretion is ubiquitous
 \citep{martinez05}. Current work suggests that while SMBHs might
spend most of their lifetime in an optically dim phase, the bulk of
mass growth occurs in the short-lived quasar stages.

The assembly of BH mass in the Universe has been tracked using optical
quasar activity. The current phenomenological approach to
understanding the assembly of SMBHs involves optical data from both
high and low redshifts. These data are used as a starting point to
construct a consistent picture that fits within the larger framework
of the growth and evolution of structure in the Universe
\citep{haehnelt98,haiman98,kauffmann00,kauffmann02,wyithe02,volonteri03,dimatteo03}.
 
Current modeling is grounded in the framework of the standard paradigm
that involves the growth of structure via gravitational amplification
of small perturbations in a CDM Universe---a model that has
independent validation, most recently from {\it Wilkinson Microwave
Anisotropy Probe} (WMAP) measurements of the anisotropies in the
cosmic microwave background \citep{spergel03,page03}. Structure
formation is tracked in cosmic time by keeping a census of the number
of collapsed dark matter halos of a given mass that form; these
provide the sites for harboring black holes. The computation of the
mass function of dark matter halos is done using either the
Press-Schechter \citep{press74} or the extended Press-Schechter theory
\citep{lacey93}, or Monte-Carlo realizations of merger trees
\citep{kauffmann00,volonteri03,bromleysomerville04} or, in some cases,
directly from cosmological N-body simulations
\citep{dimatteo03,dimatteo05}.

In particular \citet{volonteri03} have presented a detailed
merger-tree based scenario to trace the growth of black holes from the
earliest epochs to the present day. Monte-Carlo merger trees are
created for present day halos and propagated back in time to a
redshift of $\sim$ 20. With the merging history thus determined, the
initial halos at $z \sim 20$ are then populated with seed black holes
which are assumed to be remnants of the first stars that form in the
Universe. The masses of these so-called Population III stars are not
accurately known, however numerical simulations by various groups
\citep{abel00,bromm02} suggest that they are skewed to high masses of
the order of a few hundred solar masses. Seeded with the end products
of this first population, the merger sequence is followed and black
holes are assumed to grow with every major merger episode. An
accretion episode is assumed to occur as a consequence of every merger
event. Following the growth and mass assembly of these black holes, it
is required that the model is in consonance with the observed local
$M_{\rm BH} - \sigma$ relation. The luminosity function of quasars is
predicted by these models and can be compared to
observations. Volonteri et al. find that not every halo at high
redshift needs to be populated with a black hole seed in order to
satisfy the observational constraints at $z = 0$. These models do not
automatically reproduce the required abundance of supermassive black
holes inferred to power the observed $z > 6$ SDSS quasars. In order to
match the observation and produce SMBHs roughly 1 Gyr after the Big
Bang, it is required that black holes undergo brief, but extremely
strong growth episodes during which the accretion rate onto them is
well in excess of the Eddington rate \citep{volonteri05b,begelman06}.
It is the existence of these SMBHs powering quasars at $z > 6$ that
has prompted work on alternate channels to explain their mass
build-up.

In order to alleviate the problem of explaining the existence of SMBHs
in place by $z \sim 6$, roughly 1 Gyr after the Big Bang, in this
paper we examine the possibility of using a well motivated high
redshift seed black hole mass function as the initial black hole
population at the highest redshifts. We investigate the effect of
populating early dark matter halos with massive black hole seeds
predicted in a model proposed by \citet{LN06,LN07}. This model
predicts a mass function for black holes that results from the direct
collapse of pre-galactic gas discs. We study the implications of the
use of this seed mass function versus that of the Population III
remnants, in particular the difference in predictions at $z = 0$ for
the massive seed models. In Section 2, we briefly outline the high
redshift BH seed formation model, in Section 3 we evolve this model
with redshift using the merger-tree formalism developed by
\citet{volonteri03}. The results are presented in Section 4, followed
by a discussion of implications in the final section.

\section{BH seed formation model}

In this paper, we track the formation of seed black holes in an
ab-initio model and follow their mass assembly down to $z = 0$. This
is done in two separate phases - starting with the high redshift seeds
and tracing their subsequent growth. At high redshift ($z>15$), we
assume the intergalactic medium has not been significantly enriched by
metals, and therefore the gas cooling timescales are long. Under these
conditions, many authors \citep{koushiappas04,begelman06,LN06,LN07}
have shown that pre-galactic discs can efficiently transport matter
into their innermost regions through the development and amplification
of non-axisymmetric gravitational instabilities, often without
fragmentation and star formation taking place (see below). This is the
main seed formation phase, wherein massive seeds with $M\approx
10^5-10^6M_{\odot}$ can form. At lower redshifts, cooling becomes more
efficient and we assume that further accretion occurs via a `merger
driven scenario' described in more detail in Section 3. In this
section we provide simple, analytical estimates of the amount of mass
that we expect to be assembled in the form of massive BH seeds, based
on the above scenario, as a function of the key dark matter halo
parameters.  Here we refer in particular to the model by \citet{LN06},
who considered the evolution of pre-galactic discs by
self-consistently taking into account gravitational stability and
fragmentation, thereby providing a detailed inventory of the fate
of the gas.

Consider a dark matter halo of mass $M$ and virial temperature $T_{\rm
vir}$, containing gas mass $M_{\rm gas}=m_{\rm d}M$ (we also assume
that the baryon fraction is roughly 5\% implying $m_{\rm d}=0.05$), 
of primordial composition, i.e. gas not enriched 
by metals, for which the cooling function is dominated by hydrogen.  
The other main parameter characterizing a dark matter halo that is relevant to
the fate of the gas is its spin parameter $\lambda$ ($\equiv J_h E_h^{1/2}/
GM_h^{5/2}$, where $J_h$ is the total angular momentum and $E_h$ is
the binding energy). The distribution of spin parameters for dark matter halos 
measured in numerical simulations is well fit by a lognormal distribution in
$\lambda_{\rm spin}$, with mean $\bar \lambda_{\rm spin}=0.05$ and
standard deviation $\sigma_\lambda=0.5$:
\begin{equation} 
\label{plambda}
p(\lambda) \,d\lambda
={1\over \sqrt{2\pi} \sigma_\lambda}
\exp \left[-{\ln ^2 (\lambda/{\bar \lambda})
\over 2 \sigma_\lambda^2}\right] {d\lambda \over \lambda},
\end{equation}
This function has been shown to provide a good fit to the N-body results of
several investigations (e.g., \citealt{warren92,cole96,bullock01,bosch02}).

If the virial temperature of the halo $T_{\rm vir}>T_{\rm gas}$, the
gas collapses and forms a rotationally supported disc. For low values
of the spin parameter $\lambda$ the resulting disc can be compact and
dense and is subject to gravitational instabilities. This occurs when
the stability parameter $Q$ defined below approaches unity:
\begin{equation}
Q=\frac{c_{\rm s}\kappa}{\pi G \Sigma}=\sqrt{2}\frac{c_{\rm s}V_{\rm
h}}{\pi G\Sigma R},
\label{Q}
\end{equation}
where $R$ is the cylindrical radial coordinate, $\Sigma$ is the 
surface mass density, $c_{\rm s}$ is the sound speed, 
$\kappa=\sqrt{2}V_{\rm h}/R$ is the epicyclic frequency, and $V_{\rm h}$ 
is the circular velocity of the disc (mostly determined by the 
dark matter gravitational potential). We have also assumed that at the 
relevant radii ($\approx 10^2-10^3$ pc) the rotation curve is well 
described by a flat $V_{\rm h}$ profile.  We consider here the 
earliest generations of gas discs, which are of pristine composition 
with no metals and therefore can cool only via hydrogen. In thermal 
equilibrium, if the formation of molecular hydrogen is suppressed, 
these discs are expected to be nearly isothermal at a temperature of a 
few thousand Kelvin (here we take $T_{\rm gas}\approx 5000$K,
\citealt{LN06}). However, molecular hydrogen if present can cool these 
discs further down to temperatures of a few hundred Kelvin. The 
stability parameter has a critical value $Q_{c}$ of the order of 
unity, below which the disc is unstable leading to the potential 
formation of a seed black hole. The actual value of $Q_{\rm c}$ 
essentially determines how stable the disc is, with lower $Q_{\rm c}$'s 
implying more stable discs. It is well known since \citet{toomre64} 
proposed this stability criterion, that for an infinitesimally thin 
disc to be stable to fragmentation, $Q_{\rm c}=1$ for axisymmetric 
disturbances. The exact value of $Q_{\rm c}$ under more realistic 
conditions is not well determined. Finite thickness effects tend to 
stabilize the disc (reducing $Q_{\rm c}$), while on the other hand 
non-axisymmetric perturbations are in reality more unstable (enhancing 
$Q_{\rm c}$). Global, three-dimensional simulations of Keplerian discs
\citep{LR04,LR05} have shown that such discs settle down in a
quasi-equilibrium configuration with $Q$ remarkably close to unity,
implying that the critical value $Q_{\rm c}\approx 1$. In this paper,
we take $Q_{\rm c}$ to be a free parameter and evaluate our results
for a range of values. 

If the disc becomes unstable it develops non-axisymmetric spiral
structures, which leads to an effective redistribution of angular
momentum, thus feeding a growing seed black hole in the center.  This
process stops when the amount of mass transported to the center,
$M_{\rm BH}$, is enough to make the disc marginally stable. This can
be computed easily from the stability criterion in eqn.~(\ref{Q}) and
from the disc properties, determined from the dark matter halo mass
and angular momentum \citep{mo98}. In this way we obtain that the mass
accumulated in the center of the halo is given by:
\begin{equation}
M_{\rm BH}= \left\{\begin{array}{ll}
\displaystyle m_{\rm d}M\left[1-\sqrt{\frac{8\lambda}{m_{\rm d}Q_{\rm c}}
\left(\frac{j_{\rm d}}{m_{\rm d}}\right)\left(\frac{T_{\rm gas}}{T_{\rm
vir}}\right)^{1/2}}\right] & \lambda<\lambda_{\rm max} \\
0 & \lambda>\lambda_{\rm max}
\nonumber
\end{array}
\right.
\label{mbh}
\end{equation}
where 
\begin{equation}
\lambda_{\rm max}=m_{\rm d}Q_{\rm c}/8(m_{\rm d}/j_{\rm d}) (T_{\rm
  vir}/T_{\rm gas})^{1/2}
\label{lambdamax} 
\end{equation}
is the maximum halo spin parameter for which the disc is
gravitationally unstable. Note that while $Q_{\rm c} = 1$ provides the
benchmark for the onset of instability, non-axisymmetric and global
instabilities can cause the disc to become unstable for larger values
of $Q_{\rm c}$. For this reason we investigate models with $1 < Q_{\rm
c} \leq 3$.

The process described above provides a means to transport matter from
a typical scale of a few hundred parsecs down to radii of a few AU. If
the halo-disc system already possesses a massive black hole seed from
a previous generation, then this gas can provide a large fuel
reservoir for its further growth. Note that the typical accretion
rates implied by the above model are of the order of
$0.01M_{\odot}/$yr, and are therefore sub-Eddington for seeds with
masses of the order of $10^5M_{\sun}$ or so. If, on the other hand no
black hole seed is present, then this large gas inflow can form a seed
anew. The ultimate fate of the gas in this case at the smallest scales
is more uncertain. One possibility, if the accretion rate is
sufficiently large, has been described in detail by
\citet{begelman06}. The infalling material likely forms a quasi-star,
the core of which collapses and forms a BH, while the quasi-star keeps
accreting and growing in mass at a rate which would be super-Eddington
for the central BH.  Alternatively, the gas might form a super-massive
star, which would eventually collapse and form a black hole
\citep{shapiro02}.  There are no quantitative estimates of how much
mass would ultimately end up collapsing in the hole. Thus, the black
hole seed mass estimates based on eqn.~\ref{mbh} should be considered 
as upper limits.

For large halo mass, the internal torques needed to redistribute the
excess baryonic mass become too large to be sustained by the disc,
which then undergoes fragmentation. This occurs when the virial
temperature exceeds a critical value $T_{\rm max}$, given by:
\begin{equation}
\frac{T_{\rm max}}{T_{\rm gas}}>\left(\frac{4\alpha_{\rm c}}{m_{\rm
d}}\frac{1}{1+M_{\rm BH}/m_{\rm d}M}\right)^{2/3},
\label{frag}
\end{equation}
where $\alpha_{\rm c}\approx 0.06$ is a dimensionless parameter measuring the
critical gravitational torque above which the disc fragments \citep{RLA05}.

To summarize, every dark matter halo is characterized by its mass $M$
(or virial temperature $T_{\rm vir}$) and by its spin parameter
$\lambda$. The gas has a temperature $T_{\rm gas}=5000$K. If
$\lambda<\lambda_{\rm max}$ (see eqn.~\ref{lambdamax}) and $T_{\rm
vir}<T_{\rm max}$ (eqn.~\ref{frag}), then we assume that a seed BH of
mass $M_{\rm BH}$ given by eqn.~(\ref{mbh}) forms in the center. The
remaining relevant parameters are $m_{\rm d}=0.05$, $\alpha_{\rm
c}=0.06$, and we consider three different values for $Q_{\rm c}=1.5, 2,
3$, which will be referred to as (i) model A; (ii) model B and (iii)
model C which correspond respectively to cases of increasing instability
and therefore increasing efficiency for the formation of seed black
holes. What we investigate in this paper are possible
constraints/insights that the measured mass function of supermassive
black holes at $z = 0$ can provide on the onset of instability and
therefore on the efficiency of seed formation at extremely high redshifts.

\begin{figure}   
  \includegraphics[width=8cm]{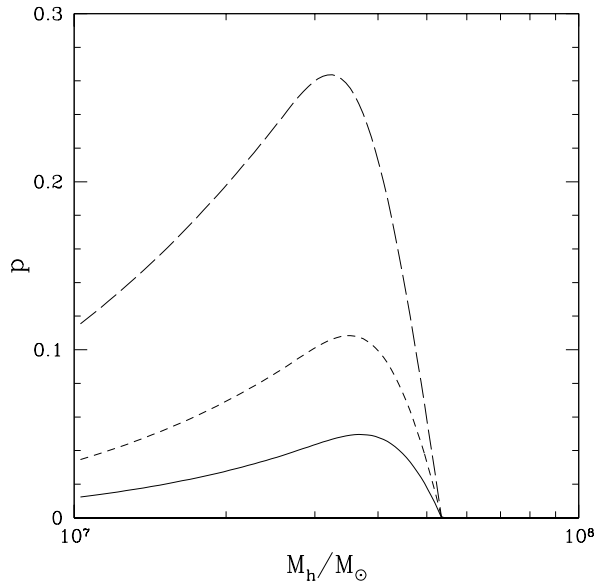}
%      	\vspace{0.5truecm}
  \caption{The probability of hosting a BH seed of any mass at $z=18$
    as a function of dark matter halo mass. The three curves refer to $Q_{\rm c}=1.5$
    (low efficiency case, solid line), $Q_{\rm c}=2$ (intermediate efficiency,
    short-dashed line) and $Q_{\rm c}=3$ (high efficiency, long-dashed line).}
   \label{fig1}
\end{figure}

To give an idea of the efficiency of BH seed formation at high $z$
within the present model, we plot in Fig.~\ref{fig1}, as an
example, the probability of forming a BH (of any mass) at $z=18$, as a
function of halo mass, for the three models. It can be seen that
typically up to 10\% of the haloes in the right mass range can form a
central seed BH, the percentage rising to a maximum of $\approx$25\%
for the high efficiency model C (highly unstable discs), and dropping
to a maximum of $\approx 4$\% for the high stability and therefore low
efficiency case (model A).

\section{The evolution of seed black holes}

We follow the evolution of the MBH population resulting from the seed
formation process delineated above in a $\Lambda$CDM Universe. Our
approach is similar to the one described in Volonteri, Haardt \&
Madau (2003). We simulate the merger history of present-day halos with
masses in the range $10^{11}<M<10^{15}\,\msun$ starting from $z=20$,
via a Monte Carlo algorithm based on the extended Press-Schechter
formalism. 

Every halo entering the merger tree is assigned a spin parameter
according to eqn.~\ref{plambda}. Recent work on the fate of halo
spins during mergers in cosmological simulations has led to
conflicting results: Vitvitska et al.  (2002) suggest that the spin
parameter of a halo increases after a major merger, and the angular
momentum decreases after a long series of minor mergers; D'Onghia \&
Navarro (2007) find instead no significant correlation between spin
and merger history. Given the unsettled nature of this matter, we
adopt Occam's razor to guide us, and assume that the spin parameter of
a halo is not modified by its merger history.

When a halo enters the merger tree we assign seed MBHs by determining
if the halo meets all the requirements described in Section 2 for the
formation of a central mass concentration. As we do not
self-consistently trace the metal enrichment of the intergalactic
medium, we consider here a sharp transition threshold, and assume that
the MBH formation scenario suggested by Lodato \& Natarajan ceases at
$z\approx 15$ \citep[see also][]{sesana07, volonteriprep2007}. 
At $z>15$, therefore, whenever a new halo appears in the merger tree 
(because its mass is larger than the mass resolution), or a pre-existing 
halo modifies its mass by a merger, we evaluate if the gaseous component 
meets the conditions for efficient transport of angular momentum to create a
large inflow of gas which can either form a MBH seed, or feed one if
already present. 

\begin{figure}   
   \includegraphics[width=8cm]{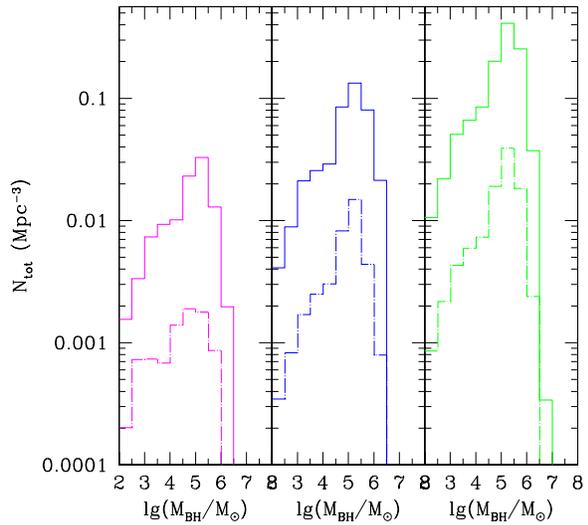} 
%      	\vspace{0.5truecm}
   \caption{Mass function of MBH seeds in three Q-models of that
differ in seed formation efficiency. Left panel: $Q_{\rm c}=1.5$ (the
least efficient model A), middle panel: $Q_{\rm c}=2$ (intermediate
efficiency model B), right panel: $Q_{\rm c}=3$ (highly efficient
model C). Seeds form at $z>15$ and this channel ceases at $z =
15$. The solid histograms show the total mass function of seeds formed
by $z=15$, while the dashed histograms refer to seeds formed at a
specific redshift, $z=18$.}
   \label{fig2}
\end{figure}

The efficiency of MBH formation is strongly dependent on critical value of 
the Toomre parameter $Q_{\rm c}$, which sets the frequency of formation, and
consequently the number density of MBH seeds. We investigate the
influence of this parameter in the determination of the global
evolution of the MBH population. Fig.~ \ref{fig2} shows the number
density of seeds formed in three models, with $Q_{\rm c}=1.5$ (low
efficiency model A), $Q_{\rm c}=2$ (intermediate efficiency model B),
and $Q_{\rm c}=3$ (high efficiency model C). The solid histograms show
the total mass function of seeds formed by $z=15$ when this formation
channel ceases, while the dashed histograms refer to seeds formed at a
specific redshift slice at $z=18$.  The number of seeds changes by
about one order of magnitude from the least efficient to the most
efficient model, consistent with the probabilities shown in Fig.~1.

We assume that, after seed formation ceases, the $z<15$ population of
MBHs evolves according to a ``merger driven scenario", as described in
\citet{volonteri06}. We assume that during major mergers MBHs accrete
gas mass that scales with the fifth power of the circular velocity
(or equivalently the velocity dispersion $\sigma_c$) of the host halo
\citep{ferrarese02}. We thus set the final mass of the MBH at the end of the
accretion episode to 90\% of the mass predicted by the $M_{\rm BH}-\sigma_c$
correlation, assuming that the scaling does not evolve with
redshift. Major mergers are defined as mergers between two dark matter
halos with mass ratio between 1 and 10. BH mergers contribute to the
mass increase of the remaining 10\%.

In order to calculate the accreting black hole luminosity function and
to follow the black hole mass growth during each accretion event, we
also need to calculate the rate at which the mass, as estimated above,
is accreted. This is assumed to scale with the Eddington rate for the
MBH, and is based on the results of merger simulations, which 
heuristically track accretion onto a central MBH
\citep{dimatteo05,hopkins05}.  The time spent by a given simulated AGN 
at a given bolometric luminosity\footnote{We convert accretion rate 
into luminosity assuming that the radiative efficiency equals the 
binding energy per unit mass of a particle in the last stable circular 
orbit. We associate the location of the last stable circular orbit to 
the spin of the MBHs, by self-consistently tracking the evolution of 
black hole spins throughout our calculations 
\citep{volonteri05}. We set 20\% as the maximum value of the 
radiative efficiency, corresponding to a spin slightly below the 
theoretical limit for thin disc accretion
\citep{thorne74}. } per logarithmic interval is approximated by
\citep{hopkins05b} as:
\begin{equation}
\label{eq:dtdlogL }
\frac{{d}t}{{d}L}=|\alpha|t_Q\, L^{-1}\, \left(\frac{L}{10^9L_\odot}\right)^\alpha,
\end{equation}
where $t_Q\simeq10^9$ yr, and $\alpha=-0.95+0.32\log(L_{\rm
peak}/10^{12} L_\odot)$. Here $L_{\rm peak}$ is the luminosity of the
AGN at the peak of its activity.  Hopkins et al. (2006) show that
approximating $L_{\rm peak}$ with the Eddington luminosity of the MBH
at its final mass (i.e., when it sits on the $M_{\rm BH}-\sigma_c$
relation) compared to computing the peak luminosity with eqn.~(6)
above gives the same result and in fact, the difference between these
2 cases is negligible. Volonteri et al. (2006) derive the following
simple differential equation to express the instantaneous accretion
rate ($f_{\rm Edd}$,in units of the Eddington rate) for a MBH of mass
$M_{\rm BH}$ in a galaxy with velocity dispersion $\sigma_c$:
\begin{equation}
\label{eq:doteddratio }
\frac{{d}f_{\rm Edd}(t)}{{d}t}=\frac{ f_{\rm Edd}^{1-\alpha}(t) }{|\alpha|
  t_Q} \left(\frac{\epsilon \dot{M}_{\rm Edd} c^2}{10^9L_\odot}\right)^{-\alpha},
\end{equation}
where here $t$ is the time elapsed from the beginning of the accretion event.
Solving this equation gives us the instantaneous Eddington ratio for a
given MBH at a specific time, and therefore we can self-consistently grow
the MBH mass. We set the Eddington ratio $f_{\rm Edd}=10^{-3}$ at
$t=0$. This same type of accretion is assumed to occur, at $z>15$,
following a major merger in which a MBH is not fed by disc
instabilities. 

In a hierarchical Universe, where galaxies grow by mergers, MBH
mergers are a natural consequence, and we trace their contribution to
the evolving MBH population (cfr. Sesana et al. 2007 for details on
the dynamical modeling). During the final phases of a MBH merger,
emission of gravitational radiation drives the orbital decay of the
binary. Recent numerical relativity simulations suggest that merging
MBH binaries might be subject to a large ``gravitational recoil": a
general-relativistic effect \citep{fitchett83,redmount89} due to the
non-zero net linear momentum carried away by gravitational waves in
the coalescence of two unequal mass black holes. Radiation recoil is a
strong field effect that depends on the lack of symmetry in the
system. For merging MBHs with high spin, in particular orbital
configurations, the recoil velocity can be as high as a few thousands
of kilometers per second 
\citep{campanelli2007a,campanelli2007b,gonzalez2007,Herrmann2007,schnittman07}. 
Here, we aim to determine the characteristic
features of the MBH population deriving from a specific seed scenario,
and its signature in present-day galaxies, we study the case without
gravitational recoil. We discuss this issue further in section 4.

\begin{figure*}   
\centerline{\epsfig{figure=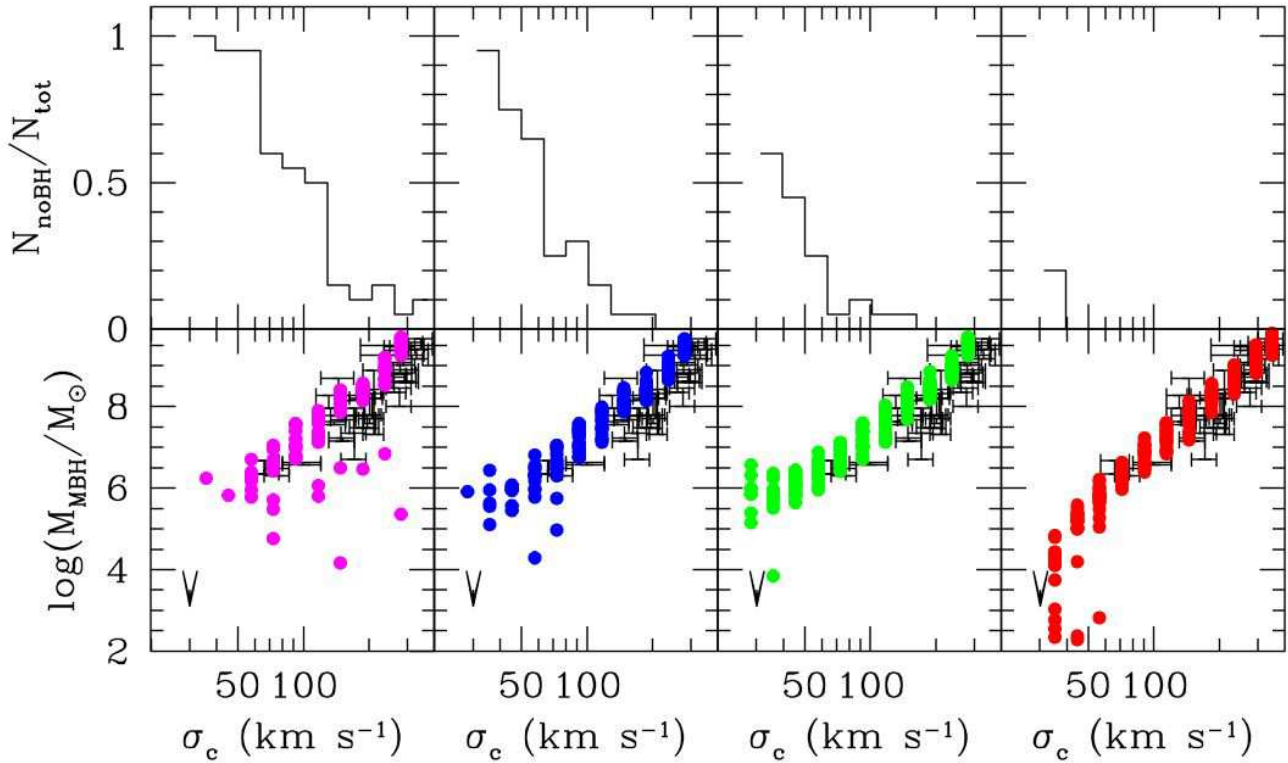,width=1.0\textwidth}}
\vspace{0.5truecm}
   \caption{ The $M_{\rm bh}-$velocity dispersion ($\sigma_c$)
     relation at $z=0$. Every circle represents the central MBH in a
     halo of given $\sigma_c$.  Observational data are marked by their
     quoted errorbars, both in $\sigma_c$, and in $M_{\rm bh}$
     (Tremaine et al. 2002).  Left to right panels: $Q_{\rm c}=1.5$,
     $Q_{\rm c}=2$, $Q_{\rm c}=3$, Population III star seeds.  {\it
     Top panels:} fraction of galaxies at a given velocity dispersion
     which {\bf do not} host a central MBH.}
   \label{fig3}
\end{figure*}

\section{Results}

Detection of gravitational waves from seeds merging at the redshift of
formation \citep{sesana07} is probably one of the best ways to
discriminate among formation mechanisms. On the other hand, the
imprint of different formation scenarios can also be sought in
observations at lower redshifts. The various seed formation scenarios
have distinct consequences for the properties of the MBH population at
$z=0$. Below, we present theoretical predictions of the various seed
models for the properties of the local SMBH population.

\subsection{Supermassive black holes in dwarf galaxies}

The repercussions of different initial efficiencies for seed formation
for the overall evolution of the MBH population stretch from
high-redshift to the local Universe. Obviously, a higher density of
MBH seeds implies a more numerous population of MBHs at later times,
which can produce observational signatures in statistical
samples. More subtly, the formation of seeds in a $\Lambda$CDM
scenario follows the cosmological bias. As a consequence, the
progenitors of massive galaxies (or clusters of galaxies) have a
higher probability of hosting MBH seeds (cfr. \citealt{madau01}). In
the case of low-bias systems, such as isolated dwarf galaxies, very
few of the high-$z$ progenitors have the deep potential wells needed
for gas retention and cooling, a prerequisite for MBH formation. We
can read off directly from Fig.~\ref{fig1} the average number of
massive progenitors required for a present day galaxy to host a
MBH. In model A, a galaxy needs of order 25 massive progenitors (mass
above $\sim10^7\msun$) to ensure a high probability of seeding within
the merger tree. In model C, instead, the requirement drops to 4
massive progenitors, increasing the probability of MBH formation in
lower bias halos.

The signature of the efficiency of the formation of MBH seeds will
consequently be stronger in isolated dwarf galaxies. Fig. \ref{fig3}
(bottom panel) shows a comparison between the observed $M_{\rm
BH}-\sigma$ relation and the one predicted by our models (shown with
circles), and in particular, from left to right, the three models
based on the \citet{LN06,LN07} seed masses with $Q_{\rm c}=1.5$, 2 and
3, and a fourth model based on lower-mass Population III star
seeds. The upper panel of Fig. \ref{fig3} shows the fraction
of galaxies that {\bf do not} host any massive black holes for different
velocity dispersion bins. This shows that the fraction of galaxies
without a MBH increases with decreasing halo masses at $z = 0$. 
A larger fraction of low mass halos are devoid of central black holes for
lower seed formation efficiencies. Note that this is one of the key
discriminants between our models and those seeded with Population III
remnants. As shown in Fig.~3, there are practically no galaxies without
central BHs for the Population III seeds.

It is interesting to note that our model predictions are in very good
agreement with the recent HST ACS census of black holes in low mass
galaxies in the Virgo cluster. \cite{ferrarese06,Wehner2006} suggest that 
below a transition galaxy mass ($\simeq 10^{10}\msun$) a central 
massive black hole seems to be replaced by a nuclear star cluster. 
Although no definite proof that Virgo dwarfs are indeed MBH-less, the 
above results imply that MBHs are more common in large galactic systems. 
Our models also indicate that a minimum velocity dispersion exists, below 
which the probability of finding a central object is very low. 

We make quantitative predictions for the local occupation fraction of MBHs. 
Our model A predicts that below $\sigma_c\approx 60\,{\rm kms}^{-1}$ the
probability of a galaxy hosting a MBH is negligible. With increasing
MBH formation efficiencies, the minimum mass for a galaxy that hosts a
MBH decreases, and it drops below our simulation limits for model
C. On the other hand, models based on lower mass Population III star
remnant seeds, predict that massive black holes might be present even
in low mass galaxies.

We note here that in our investigation we have not included any
mechanism that could further lower the occupation fraction of MBHs
(e.g., gravitational recoil, three-body MBH interactions). For any
value of $Q_{\rm c}$ the occupation fraction computed above is
therefore an upper limit.

Although there are degeneracies in our modeling (e.g., between the
minimum redshift for BH formation and instability
criterion), the BH occupation fraction, and the masses of the BHs in
dwarf galaxies are the key diagnostics. In local observations, the
clearest signatures of massive seeds compared to Population III
remnants, would be a lower limit of order the typical mass of seeds
(Fig. ~\ref{fig2}) to the mass of MBHs in galaxy centers, as shown in
Fig. ~\ref{fig3}. An additional caveat worth mentioning is the
possibility that a galaxy is devoid of a central MBH because of
dynamical ejections (due to either the gravitational recoil or
three-body scattering). The signatures of such dynamical interactions
should be more prominent in dwarf galaxies, but ejected MBHs would
leave observational signatures on their hosts (G{\"u}ltekin et al. in
prep.). On top of that, \citet{schnittman07} and \citet{volonteri07}
agree in considering the recoil a minor correction to the overall
distribution of the MBH population at low redshift (cfr. figure 4 in
Volonteri 2007).

Additionally, as MBH seed formation requires halos with low angular
momentum (low spin parameter), we envisage that low surface
brightness, bulge-less galaxies with high spin parameters (i.e. large
discs) are systems where MBH seed formation is less
probable. \footnote{This prediction is pertinent to all models relying
on gravitational instabilities triggered in low spin parameter halos.}
Furthermore, bulgeless galaxies are believed to preferentially have
quieter merger histories and are unlikely to have experienced any
major merger, which could have brought in a MBH from a companion
galaxy. The (possible) absence of a MBH in M33 hence arises naturally
(Merritt, Ferrarese \& Joseph 2001; Gebhardt et al. 2001) in our
model.

 \begin{figure}   
   \includegraphics[width=8cm]{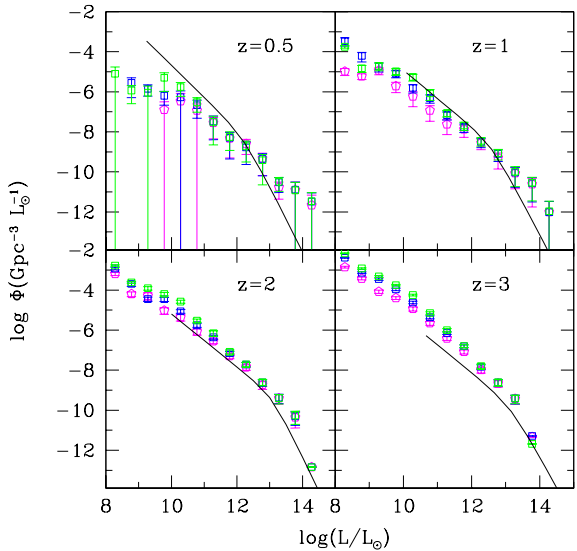} 
%      	\vspace{0.5truecm}
   \caption{Bolometric luminosity functions at different
redshifts. All 3 models match the observed bright end of the LF at
high redshifts and predict a steep slope at the faint end down to $z =
1$. The 3 models are not really distinguishable with the LF. However
at low redshifts, for instance at $z = 0.5$, all 3 models are
significantly flatter at both high and low luminosities and do not
adequately match the current data. As discussed in the text, the LF is
strongly determined by the accretion prescription and what we see here
is simply a reflection of that fact.}
   \label{fig4}
\end{figure}

\subsection{The luminosity function of accreting black holes}

Turning to the global properties of the MBH population, as suggested
by Yu \& Tremaine (2002) the mass growth of the MBH population at
$z<3$ is dominated by the mass accreted during the bright epoch of
quasars, thus washing out most of the imprint of initial conditions.
This is evident when we compute the luminosity function of AGN.
Clearly the detailed shape of the predicted luminosity function
depends most strongly on the accretion prescription used. With our
assumption that the gas mass accreted during each merger episode is
proportional to $V_c^5$, we find that distinguishing between the
various seed models is difficult. As shown in Fig.~\ref{fig4}, all 3
models reproduce the bright end of the observed bolometric LF
\citep{hopkins07} at higher redshifts (marked as the solid curve in
all the panels), and predict a fairly steep faint end that is as yet
undetected. All models fare less well at low redshift, shown in
particular at $z = 0.5$. This could be due to the fact that we have
used a single accretion prescription to model growth through
epochs. On the other hand, the decline in the available gas budget at
low redshifts (since the bulk of the gas has been consumed by this
epoch by star formation activity) likely changes the radiative
efficiency of these systems. Besides, observations suggest a sharp
decline in the number of actively accreting black holes at low
redshifts across wave-lengths, produced most probably due to changes
in the accretion flow as a result of change in the geometry of the
nuclear regions of galaxies. In fact, all 3 of our models
under-predict the slope at the faint end. There are three other effects
that could cause this flattening of the LF at the faint end at low
redshift for our models: (i) not having taken into account the fate of
on-going merging and the fate of satellite galaxies; (ii) the number
of realizations generated and tracked is insufficient for statistics,
as evidenced by the systematically larger errorbars and (iii) more
importantly, it is unclear if merger-driven accretion is indeed the
trigger of BH fueling in the low redshift universe. We note that the 3
massive seed models and Population III seed model cannot be
discriminated by the LF at high redshifts. Models B and C are also in
agreement viz-a-viz the predicted BH mass function at $z = 6$ (see Fig.~2), even
assuming a very high radiative efficiency (up to 20\%), while model A might
need less severe assumptions, in particular for BH masses larger than
$10^7\,\msun$.

\subsection{Comoving mass density of black holes}
 
\begin{figure}   
   \includegraphics[width=8cm]{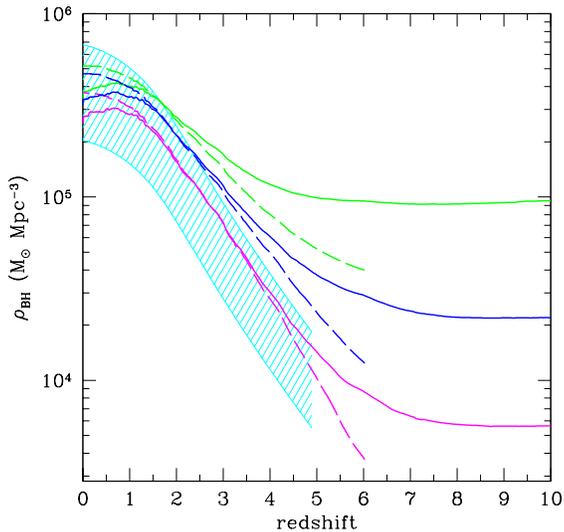} 
%      	\vspace{0.5truecm}
   \caption{Integrated black hole mass density as a function of
     redshift. Solid lines: total mass density locked into nuclear
     black holes.  Dashed lines: integrated mass density {\rm
     accreted} by black holes.  Models based on BH remnants of
     Population III stars (lowest curve), $Q_{\rm c}=1.5$ (middle
     lower curve), $Q_{\rm c}=2$ (middle upper curve), $Q_{\rm c}=3$
     (upper curve).  Shaded area: constraints from So{\l}tan-type
     arguments, where we have varied the radiative efficiency from a
     lower limit of 6\% (applicable to Schwarzschild MBHs, upper
     envelope of the shaded area), to about 20\% (Wang et
     al. 2006). All 3 massive seed formation models are in comfortable
     agreement with the mass density obtained from integrating the
     optical luminosity functions of quasars.}
   \label{fig5}
\end{figure}

Since during the quasar epoch MBHs increase their mass by a large factor,
signatures of the seed formation mechanisms are likely more evident at
{\it earlier epochs}. We compare in Fig.~\ref{fig5} the integrated
comoving mass density in MBHs to the expectations from So{\l}tan-type
arguments (F. Haardt, private communication), assuming that quasars
are powered by radiatively efficient flows (for details, see
\citealt{yu02,elvis02,marconi04}). While during and after the quasar
epoch the mass densities in models A, B, and C differ by less than a
factor of 2, at $z>3$ the differences become more pronounced.

A very efficient seed MBH formation scenario can lead to a very large
BH density at high redshifts. For instance, the highest efficiency
model C with $Q_{\rm c}=3$, the integrated MBH density at $z=10$ is
already $\sim 25\%$ of the density at $z=0$. The plateau at $z>6$ is
due to our choice of scaling the accreted mass with the $z=0$ $M_{\rm
bh}-$velocity dispersion relation. Since in our models we let MBHs
accrete a mass which scales with the fifth power of the circular
velocity of the halo, the accreted mass is a small fraction of the MBH
mass (see the discussion in \citealt{marulli06}), and the overall
growth remains small, as long as the mass of the seed is larger than
the accreted mass, which, in our assumed scaling, happens when the
mass of the halo is below a few times $10^{10}\msun$. The comoving
mass density, an integral constraint, is reasonably well determined
out to $z = 3$ but is poorly known at higher redshifts. All models
appear to be satisfactory and consistent with current observational
limits (shown as the shaded area).

\subsection{Black hole mass function at $z=0$}

\begin{figure}   
   \includegraphics[width=8cm]{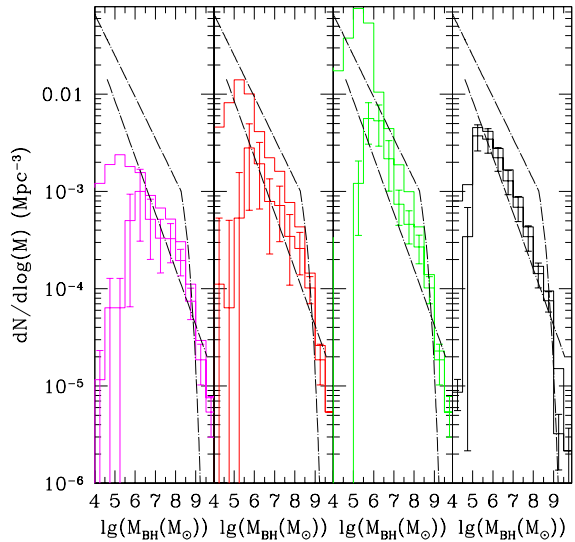} 
   \caption{Mass function of black holes at z=0. Histograms represent
     the results of our models, including central galaxies only (lower
     histograms with errorbars), or including satellites in groups and
     clusters (upper histograms). Left panel: $Q_{\rm c}=1.5$, mid-left
     panel: $Q_{\rm c}=2$, mid-right panel: $Q_{\rm c}=3$, right panel:
     models based on BH remnants of Population III stars.  
     Upper dashed line: mass function derived from combining the velocity
     dispersion function of Sloan galaxies (Sheth et al. 2003, where
     we have included the late-type galaxies extrapolation), and BH
     mass-velocity dispersion correlation (e.g., Tremaine et
     al. 2002). Lower dashed line: mass function derived using the
     Press-Schechter formalism from Jenkins et al. (2001) in
     conjunction with the $M_{\rm BH} -\sigma$ relation (Ferrarese
     2002).}
   \label{fig6}
\end{figure}

One of the key diagnostics is the comparison of the measured and
predicted BH mass function at $z = 0$ for our 3 models. In
Fig.~\ref{fig6}, we show (from left to right, respectively) the mass
function predicted by models A, B, C and Population III remnant seeds
compared to that obtained from measurements.  The histograms show the
mass function obtained with our models (where the upper histogram
includes all the black holes while the lower one only includes black
holes found in central galaxies of halos in the merger-tree
approach). The two lines are two different estimates of the observed
black hole mass function. In the upper one, the measured velocity
dispersion function for nearby late and early-type galaxies from the
SDSS survey \citep{bernardi03,sheth03} has been convolved with the
measured $M_{\rm BH} - \sigma$ relation. We note here that the scatter
in the $M_{\rm bh} - \sigma$ relation is not explicitly included in
this treatment, however the inclusion of the scatter is likely to
preferentially affect the high mass end of the BHMF, which provides
stronger constraints on the accretion histories rather than the seed
masses. It has been argued \citep[e.g.,][]{Tundo2007,Bernardi2007,Laueretal2007} 
that the BH mass function differs if the bulge mass is used instead of the 
velocity dispersion in relating the BH mass to the host galaxy. 
Since our models do not trace the formation and growth of stellar bulges 
in detail, we are restricted to using the velocity dispersion in our analysis. 

%On the other hand, if the $M_{\rm bh} - M_{\rm bulge}$
%relation is used to determine the BHMF at $z = 0$ since it has more
%scatter than the $M_{\rm bh} - \sigma$ relation, there is more
%resulting uncertainty.

The lower dashed curve is an alternate theoretical estimate of the BH mass 
function derived using the Press-Schechter formalism from  \citet{jenkins01} 
in conjunction with the observed $M_{\rm BH} -\sigma$ relation. Selecting only the
central galaxies of halos in the merger-tree approach adopted here
(lower histograms) is shown to be fairly equivalent to this analytical
estimate, and this is clearly borne out as is evident from the plot.
When we include black holes in satellite galaxies (upper histograms,
cfr. the discussion in Volonteri, Haardt \& Madau 2003) the predicted
mass function moves towards the estimate based on SDSS galaxies. The
higher efficiency models clearly produce more BHs. At higher
redshifts, for instance at $z = 6$, the mass functions of active MBHs
predicted by all models are in very good agreements in particular for
BH masses larger than $10^6\,\msun$, as it is the growth by accretion
that dominates the evolution of the population. At the highest mass
end ($>10^9\,\msun$) model A lags behind models B and C, although we
stress once again that our assumptions for the accretion process are
very conservative.

The {\it relative} differences between models A, B, and C at the
low-mass end of the mass function, however, are genuinely related to
the MBH seeding mechanism (see also Figs.~\ref{fig3} and
~\ref{fig5}). In model A, simply, fewer galaxies host a MBH, hence
reducing the overall number density of black holes. Although our
simplified treatment does not allow robust quantitative predictions,
the presence of a "bump" at $z = 0$ in the MBH mass function at the
characteristic mass that marks the peak of the seed mass function
(cfr. Fig.~\ref{fig2}) is a sign of highly efficient formation of
massive seeds (i.e., much larger mass with respect, for instance,
Population III remnants). The higher the efficiency of seed formation,
the more pronounced is the bump (note that the bump is most prominent
for model C). Since current measurements of MBH masses extend barely
down to $M_{\rm bh}\sim 10^6 \msun$, this feature cannot be
observationally tested with present data but future campaigns, with
the Giant Magellan Telescope or JWST, are likely to extend the mass
function measurements to much lower black hole masses.

\section{Discussion and implications}

In this paper, we have investigated the role that the choice of the
initial seed black hole mass function at high redshift ($z \sim 18$)
plays in the determination of observed properties of local quiescent
SMBHs. While the errors on mass determinations of local black holes
are large at the present time, definite trends with host galaxy
properties are observed. The tighest correlation appears to be between
the BH mass and the velocity dispersion of the host spheroid. Starting
with the ab-initio black hole seed mass function computed in the
context of direct formation of central objects from the collapse of
pre-galactic discs in high redshift halos, we follow the assembly
history to late times using a Monte-Carlo merger tree approach. Key to
our calculation of the evolution and build-up of mass is the
prescription that we adopt for determining the precise mass gain
during a merger. Motivated by the phenomenological observation of
$M_{\rm BH} \propto V_{\rm c}^5$, we assume that this proportionality
carries over to the gas mass accreted in each step. With these
prescriptions, a range of predictions can be made for the mass
function of black holes at high and low $z$, and the integrated mass
density of black holes, all of which are observationally
determined. We evolve 3 models, designated model A, B and C which
correspond to increasing efficiencies respectively for the formation
of seeds at high redshift. These models are compared to one in which
the seeds are remnants of Population III stars. 

It is important to
note here that one major uncertainty prevents us from making more
concrete predictions: the unknown metal enrichment history of
the Universe. Key to the implementation of our models is the choice of
redshift at which massive seed formation is quenched. The direct
seed formation channel described here ceases to operate once the Universe
has been enriched by metals that have been synthesized after the 
first generation of stars have gone supernova. Once metals are available
in the Inter-Galactic Medium gas cooling is much more efficient and hydrogen
in either atomic or molecular form is no longer the key player. In this 
work we have assumed this transition redshift to be $z = 15$. The efficiency 
of MBH formation and the transition redshift are somehow degenerate (e.g., 
a model with $Q=1.5$ and enrichment redshift $z=12$ is halfway between 
model A and model B); if other constraints on this redshift were available we 
could considerably tighten our predictions. 
 
Below we list our predictions and compare how they fare with respect
to current observations. The models investigated here clearly differ
in predictions at the low mass end of the black hole mass function. 
With future observational sensitivity in this domain, these models 
can be distinguished.

Our model for the formation of relatively high-mass black hole seeds
in high-$z$ halos has direct influence on the black hole occupation
fraction in galaxies at $z=0$. This effect is more pronounced for low
mass galaxies.  We find that a significant fraction of low-mass
galaxies might not host a nuclear black hole. This is in very good
agreement with the shape of the $M_{\rm bh} - \sigma$ relation
determined recently from an observational census (an HST ACS survey)
of low mass galaxies in the Virgo cluster reported by Ferrarese et
al. (2006).

 The models studied here (with different black hole seed formation
  efficiency) are distinguishable at the low mass end of the BH mass
  function, while at the high mass end the effect of initial seeds
  appears to be sub-dominant. While current data in the low mass
  regime is scant \citep{Barth2004,Greene2007}, future instruments and surveys 
  are likely to probe this region of parameter space with significantly higher
  sensitivity.

All our models predict that low surface brightness, bulge-less
  galaxies with high spin parameters (i.e. large discs) are systems
  where MBH formation is least probable.
  
  One of the key caveats of our picture is that it is unclear if the
  differences produced by different seed models on observables at 
  $z = 0$ might be compensated or masked by BH fueling modes at earlier
  epochs. There could be other channels for BH growth that dominate at
  low redshifts like minor mergers, dynamical instabilities, accretion of
  molecular clouds, tidal disruption of stars. The decreased importance of the
  merger driven scenario is patent from observations of low-redshift AGN, 
  which are for the large majority hosted by undisturbed galaxies 
  \citep[e.g.,][and references therein] {Pierce2007} in 
  low-density environments \citep[e.g.,][]{Li2006}. However, the 
  feasibility and efficiency of some alternative channels are still to 
  be proven (for example, about the efficiency of feeding from large scale 
  instabilities see discussion in \citealt{KingPringle2007,shlosman89,goodman03,collin99}). 
  In any event, while these additional channels for BH {\it growth} can modify the 
  detailed shape of the mass function of MBHs, or of the luminosity function of 
  quasars, they will not create a new MBH.  The occupation fraction 
  of MBHs (see figure 3) is therefore largely {\it independent} of the 
  accretion mechanism and a true signature of the formation process.
    
 To date, most theoretical models for the evolution of MBHs in galaxies
do not include {\it how} MBHs form. This work is a first analysis of
the observational signatures of massive black hole formation
mechanisms in the low redshift universe, complementary to the
investigation by Sesana et al. (2007), where the focus was on
detection of seeds at the very early times where they form, via
gravitational waves emitted during MBH mergers. We focus here on
possible dynamical signatures that forming massive black hole seeds
carries over to the local Universe. We believe that the signatures of
seed formation mechanisms will be far more clear if considered jointly
with the evolution of the spheroids that they host. The mass, and
especially the frequency, of the forming MBH seeds is a necessary
input when investigating how the feedback from accretion onto MBHs
influences the host galaxy, and is generally introduced in numerical
models using extremely simplified, {\it ad hoc} prescriptions (e.g.,
\citealt{springel05,dimatteo05,hopkins06,croton05,cattaneo06,bower06}).
Adopting more detailed models for black hole seed formation, as outlined 
here, can in principle strongly affect such results.  For
instance, \citet{kauffmann04} find that AGN activity is typically
confined to galaxies with $M>10^{10}\msun$. If we consider the
occupation fraction of MBHs in such galaxies, we find that it differs
by a large factor between models A and C, being of order 10\% in the
low efficiency model (at $z\sim 1-4$) and 50\% or higher in model
C. Consequently, the possibility of AGN feedback and its effect on the
host would be selective in the former case, or widespread in the
latter case. Adopting sensible assumptions for the masses, and
frequency of MBH seeds in models of galaxy formation is necessary if
we want to understand the symbiotic growth of MBHs and their hosts.

\section*{Acknowledgments}

PN and MV acknowledge the 2006 KITP program titled `The Physics of Galactic
Nuclei', supported in part by the National Science Foundation under Grant No.
PHY99-07949.

\bibliographystyle{mn2e}
\bibliography{lodato}

\end{document}